\documentclass[aps,pra,twocolumn,tightenlines,superscriptaddress]{revtex4-1}
\usepackage{amsfonts,amssymb}
\usepackage{amsmath}
\usepackage{graphicx,times,color}
\usepackage[breaklinks=true]{hyperref}
\usepackage{bm,relsize}

\def\prx{Phys. Rev. X}
\def\){\right)}
\def\({\left(}
\def\]{\right]}
\def\[{\left[}

\newcommand{\be}{\begin{equation}}
\newcommand{\ee}{\end{equation}}

\newcommand{\roughly}[1]%
{\mathrel{\raise.4ex\hbox{$#1$\kern-.75em\lower1ex\hbox{$\sim$}}}}

\newcommand\beq{\begin{eqnarray}}\newcommand\eeq{\end{eqnarray}}

\def\Dsl{\,\raise.15ex \hbox{/}\mkern-12.8mu D}

\def\fm3{fm$^{-3}$}

\renewcommand{\mathbf}[1]{{\vec{#1}}}

\begin{document}

\title{The Gor'kov and Melik-Barkhudarov correction to the mean-field critical field transition to Fulde-Ferrell-Larkin-Ovchinnikov states}

\author{Heron Caldas} 
\email{hcaldas@ufsj.edu.br}
\affiliation{Departamento de Ci\^{e}ncias Naturais, Universidade Federal de S\~{a}o Jo\~{a}o Del Rei, Pra\c{c}a Dom Helv\'{e}cio 74, 36301-160, S\~{a}o Jo\~{a}o Del Rei, MG, Brazil}

\author{Qijin Chen}
\email{qchen@uchicago.edu}
 \affiliation{Shanghai Branch, National Laboratory for Physical Sciences at Microscale and School of Physical Sciences, University
  of Science and Technology of China, Shanghai 201315, China}

\begin{abstract}
  The Fulde-Ferrell-Larkin-Ovchinnikov (FFLO) states, characterized by
  Cooper pairs condensed at finite-momentum are, at the same time,
  exotic and elusive. It is partially due to the fact that the FFLO
  states allow superconductivity to survive even in strong magnetic
  fields at the mean-field level. The effects of induced
  interactions at zero temperature are calculated in both clean and dirty
  cases, and it is found that the critical field at which the quantum
  phase transition to an FFLO state occurs at the mean-field level is
  strongly suppressed in imbalanced Fermi gases. This strongly shrinks
  the phase space region where the FFLO state is unstable and more
  exotic ground state is to be found.  In the presence of high level
  impurities, this shrinkage may destroy the FFLO state completely.
  \hfill \color{blue}Ann. Phys. (Berlin) \textbf{2020}, 2000222; DOI: 10.1002/andp.202000222
\end{abstract}
\maketitle

\section{Introduction}

Fermionic particles with two different spins occupying states of momenta
with equal size but in opposite directions close to their common Fermi
surface form Cooper pairs, when subject to a pairing interaction. This
is successfully explained by the Bardeen-Cooper-Schrieffer (BCS)
theory of superconductivity~\cite{BCS}. The presence of an imbalance
between the two spin configurations prevents this mechanism, since
there are now two Fermi surfaces that do not coincide so that pairing
with zero total momentum for the BCS state is energetically
unfavorable, as the formation of Cooper pairs implies equal densities
of the two spin species~\cite{chen07prb,Teo1,Inguscio2007,Review2}.

The difficulty of BCS pairing caused by spin imbalance led to the
proposal of possible energetically more favorable
Fulde-Ferrell-Larkin-Ovchinnikov (FFLO) states ~\cite{FF,LO}, which
Bose condense into a finite momentum $\mathbf{q}$ or a pair of momenta
$\pm \mathbf{q}$ (for FF and LO states, respectively).
  The FF state features a single-plane-wave superfluid order
  parameter,
  $\Delta_\mathbf{R} = \Delta e^{i\mathbf{q}\cdot\mathbf{R}}$, which
  has a spatially uniform amplitude. And the LO phase has a
  standing-wave-like order parameter,
  $\Delta_\mathbf{R} = \Delta \cos(\mathbf{q}\cdot\mathbf{R})$, which
  is a superposition of two counterpropagating plane waves, and is
  inhomogeneous in both amplitude and phase. The LO phase can be
  generalized to higher order crystalline states with multiple
  plane-wave components. While a stable FFLO state may exist in an
anisotropic system \cite{ShimaharaPB,ShimaharaJPSJ67} or in a lattice
\cite{Samokhin2006PRB,ShimaharaJPSJ67,Kinnunen_2018}, especially in a
low dimensions \cite{Koponen_2008,Ptok2017,Hulet_2016PRL}, however, it
has been shown that the FFLO states are intrinsically unstable in
clean homogeneous three-dimensional (3D) and 2D
continuum systems \cite{Insta}.  Instead, noncondensed pairing with
the lowest pair energy at finite momenta is expected, which may lead
to exotic ground states.  Thus it is important to find the true
solution where the unstable mean-field FFLO solutions exist. While
this is a very difficult issue, in this paper, we aim to further
constrain the phase space region where exotic pairing state may
live. In particular, we find that particle-hole fluctuations help to
significantly reduce the critical field transition window for the
mean-field FFLO states. This also implies that the regions that
otherwise have a mean-field FFLO solution should now exhibit more
conventional solutions, such as polaronic normal phase
\cite{ZwierleinPolaron,SalomonPolaron} and phase separation
\cite{PhaseSep,Exp2}.

At the mean-field (MF) level, for small asymmetries between the two
spin species, and at zero temperature $T$, the system persists as a
BCS superfluid of zero momentum. However, when the imbalance between
the two Fermi surfaces is too large, superfluid pairing is broken
apart so that the system undergoes a quantum phase transition to the
normal state. Therefore, for a given imbalance, there exists a lower
threshold of pairing strength for the BCS pairing solution to
exist. On the other hand, for a given interaction strength, there
exists an upper bound for the imbalance before pairing is broken. The
existence of such a transition at a critical value of the polarization
was first realized by Clogston~\cite{Clogston} and
Chandrasekhar~\cite{Chandrasekhar}, who independently predicted the
occurrence of a first-order phase transition from the superfluid to
the normal state. This is known as the Clogston-Chandrasekhar (CC)
limit of superfluidity, and was originally proposed in the context of
conventional superconductivity.

Stability analysis based on energetic considerations reveals that the
mean-field BCS solution at $T=0$ is not stable in the presence of
imbalance until the system enters the Bose-Einstein condensation (BEC)
regime where the gap and hence the condensation energy become
large~\cite{Chien06,PWY05,Stability,Sheehy2006}.  Indeed, the momentum of the
minority fermions would have to be lifted up to match that of their
majority partners, but the energy cost is larger than the condensation
energy gain when the pairing gap is small. As a consequence, thermal
smearing of the Fermi surfaces leads to possible intermediate
temperature superfluidity, at both the mean-field level and with
fluctuations included \cite{Chien06}.

Theoretical investigations with ultracold imbalanced Fermi gases,
where the numbers of atoms in the two spin states are different, have
predicted that the first order transition between the superfluid at equal
spin population and the imbalanced normal mixture brings about a phase
separation between coexisting normal and superfluid
phases~\cite{Bedaque2003,Caldas2004}. Recent experiments of Fermi
gases in a trap using tomographic techniques have found a sharp
separation between a superfluid core at the trap center and a
partially polarized normal phase outside the core \cite{PhaseSep,Exp2}.
So far, the exploration of a two-component Fermi gas with imbalanced
populations remains a current and active area of research in the field
of ultracold atoms in both theory~\cite{Teo1,Teo2,Teo3,Zwerger2012} and
experiment~\cite{Exp1,Exp2,Exp3,Exp4,Exp5}, which gives to this field
the unprecedented opportunity for mimicking and simulating condensed
matter systems~\cite{Lewenstein,QI}. Particularly,  population
imbalance in atomic Fermi gases can access the full range between 0
and 100\%, making it an ideal platform for studying the FFLO physics.

The FFLO superfluid state was proposed independently by Fulde and
Ferrell \cite{FF}, and Larkin and Ovchinnikov \cite{LO}, to
address the possibility for the Cooper pairs in an $s$-wave
superconductor in the presence of a Fermi surface mismatch caused by a
Zeeman field to have a non-zero total momentum $\vec{q}$, with a
spatially modulated superfluid order parameter.  In this {\it
  intriguing} pairing mechanism, the superfluidity ``perseveres'' in
the form of an FFLO state, with a spatial modulation of the phase and
amplitude of the order parameter for the FF and LO states,
respectively.  In the last 55 years many groups have tried to find the
FFLO phase experimentally, and some have found only indirect
signatures as, for example, in the heavy fermion superconductor
CeCoIn$_5$~\cite{Pagliuso,Radovan,Kenzelmann,Kumagai,Koutroulakis,TokiwaPRL109,KimPRX6,HatakeyamaPRB}, organic superconductors \cite{Mayaffre,Koutroulakis,Varelogiannis,Lortz_2007PRL,YonezawaPRL100,BergkPRB83,ConiglioPRB,Singleton_2000,Agosta_2017PRL,Wosnitza_2017,Agosta_2018,Sugiura_2019}, as well as iron-based superconductors \cite{Ptok2013,Cho_83PRB,Cho_PRL119}. Theoretical
studies have found that the FFLO states become unstable in various
situations \cite{Marenko,Ashvin,RadzihovskyPRA84}. In particular, it
is shown in Ref.~\cite{Insta} that due to the inevitable pairing
fluctuations (including both amplitude and phase), the FFLO states are
intrinsically unstable in isotropic 3D and 2D systems, and thus exotic
pairing state may emerge.

So far, the true stable solution is still unknown, where the above
mean-field FFLO solution is stable against phase separation but
unstable due to pairing fluctuations.  As a first step, in the present
paper, we consider the contributions of particle-hole fluctuations and
study their effect on the relevant phases.

It has been known that particle-hole fluctuations may have a strong
effect in the solution of the superfluid transition temperature $T_\text{c}$
in the BCS--BEC crossover \cite{Review,Qijin}. Namely, there is
a change in the coupling of the interaction due to screening of the
interspecies (or induced) interaction, known as the Gor'kov and
Melik-Barkhudarov (GMB) correction~\cite{GMB61}. On the BEC side of
the unitarity, where the two-body scattering length diverges
\footnote{The unitary limit also corresponds to the threshold
  interaction strength at which a bound state starts to emerge in
  vacuum.}, fluctuations in the pairing channel are dominant, while
the GMB fluctuations become weaker towards the BEC side and usually is
taken as vanishing in this region due to the disappearance of the
Fermi surface.

The effect of induced interactions was first considered by Gor'kov and
Melik-Barkhudarov, who found that in a dilute 3D balanced spin-$1/2$
Fermi gas the (overestimated) MF transition temperature is suppressed
by a factor $(4e)^{1/3} \approx 2.2$~\cite{GMB61}.

Quite generally, the calculation of the GMB correction has been
restricted to the balanced case, with the Zeeman field
$h = (\mu_{\uparrow} - \mu_{\downarrow})/2=0$, where $\mu_{\uparrow}$
and $\mu_{\downarrow}$ are the chemical potential of the spin-up and
spin-down fermions, respectively. Previous investigations with
$h \neq 0$ were focused on the effects of the induced interactions on
the tricritical point $(p_\text{t},T_\text{t}/T_\text{F})$ of imbalanced Fermi gases in
3D~\cite{Yu10} and 2D~\cite{Resende}. Here $p_\text{t}$ and $T_\text{t}$ are the
polarization and the temperature at the tricritical point, and $T_\text{F}$
is the Fermi temperature. (And we define Fermi energy and Fermi
momentum via $E_\text{F} =k_\text{B}T_\text{F} = \hbar^2 k_\text{F}^2 /2m$). More sophisticated
calculations involving self-consistent feedback effect from both
particle-particle and particle-hole channels can be found in
Ref.~\cite{Qijin}.

Another important factor that has a strong impact on the superfluid
phase is disorder and impurity scattering. The effects of disorder on
an FFLO state have been investigated in $s$-
\cite{Aslamazov_1959,Takada_1970,Bulaevskii_1976,Agterberg_2001,IkedaPRB81,YangPRB2008}
and $d$-wave
\cite{Agterberg_2001,Adachi_2003PRB,Ting_2007PRB,Vorontsov_2008PRB,Yanase_2009,Datta_2019PRB}
superconductors.  While weak nonmagnetic impurities have been benign
to $s$-wave BCS superconductors a la the Anderson's theorem
\cite{Anderson}, they may suppress or destroy an FFLO state
\cite{Ptok_2010,YangPRB2008}.

In this paper, we investigate at the mean-field level the effects of
the GMB correction on the FFLO transition that may occur in Fermi
gases with imbalanced spin populations, in the clean limit and in the
presence of nonmagnetic impurities. We study the continuous phase
transition that is triggered by an increase in the chemical potential
imbalance $h$, in homogeneous 3D systems.
We find the GMB correction to the critical chemical potential
imbalance $h_\text{s}$ responsible for the phase transitions from the
partially polarized (PP) FFLO phase to a fully polarized (FP) normal
state. In the presence of high level impurities, we show that short
lifetimes necessarily further decrease $h_\text{s}$ and, consequently,
reduce or even completely destroy the predicted FFLO region of
existence. To our knowledge, this is the first time that the GMB
correction is considered in the context of FFLO physics.

The paper is organized as follows. In Sec.~\ref{N-M} we first
calculate the generalized pair susceptibility in the clean case,
associated with the onset of the instability of the PP normal
phase. In Sec.~\ref{ind} we obtain the induced interactions and find
its effects on the critical chemical potential imbalance $h_\text{s}$
which sets the transition to the FFLO phase. In Sec.~\ref{Ap} we show
how the GMB correction further reduces the FFLO window in the presence
of nonmagnetic impurities.  Finally, we conclude in Sec.~\ref{conc}.

\section{The Intermediate Normal-Mixed Phase}
\label{N-M}

We consider a generic system of fermions characterized by an
effective, short range pairing interaction $-g$, (where $g>0$), with
grand canonical Hamiltonian in momentum space \cite{Chen1,chen07prb}
\begin{eqnarray}
H & = & \sum_{\mathbf{k}\sigma} \xi_{\mathbf{k}\sigma}
c^{\dag}_{\mathbf{k}\sigma} c^{\ }_{\mathbf{k}\sigma}
\nonumber \\
  & & - g \sum_{\mathbf{k}\mathbf{k}'\mathbf{q}} 
c^{\dag}_{\mathbf{k} +\mathbf{q}/2\uparrow} 
c^{\dag}_{-\mathbf{k}+\mathbf{q}/2\downarrow} 
c^{\ }_{-\mathbf{k}'+\mathbf{q}/2\downarrow} 
c^{\ }_{\mathbf{k}'+\mathbf{q}/2\uparrow},
\label{Hamiltonian}
\end{eqnarray}
where the bare dispersion
$\xi_{\mathbf{k}\sigma} = \epsilon_\mathbf{k} -\mu_\sigma =
\mathbf{k}^2/2m -\mu_\sigma$, and $\sigma = \uparrow, \downarrow$ is
the spin index. Here $c^\dag$ ($c$) is the fermion creation
(annihilation) operator, and we have set the system volume to
unity. We shall also take the natural units, $\hbar = k_\text{B} = 1$. The
average chemical potential $\mu = (\mu_\uparrow + \mu_\downarrow)/2$.
The population imbalance is defined as the relative spin density difference, 
$p=(n_\uparrow-n_\downarrow)/(n_\uparrow+n_\downarrow)$.  At the
mean-field level, the reduced Hamiltonian for one-plane-wave FFLO
state with pairing between $\mathbf{k}$ and $-\mathbf{k}+\mathbf{q}$
states is given by \cite{He2007}
\begin{eqnarray}
H^\text{MF}&=&\sum_{\mathbf{k}} \Big\{\xi_{\mathbf{k}\uparrow}
c_{\mathbf{k}\uparrow}^{\dag}c_{\mathbf{k}\uparrow} 
+\xi_{\mathbf{k}-\mathbf{q}\downarrow}
c_{-\mathbf{k}+\mathbf{q}\downarrow}^{\dag}c_{-\mathbf{k}+\mathbf{q}\downarrow} 
 \nonumber \\
& &{}+\Delta_\mathbf{q} c_{-\mathbf{k}+\mathbf{q}\downarrow}^{\dag}
 c_{\mathbf{k}\uparrow}^{\dag}+\Delta_\mathbf{q}^*
 c_{\mathbf{k}\uparrow}c_{-\mathbf{k}+\mathbf{q}\downarrow}\Big\} \,.
\label{eq:19}
\end{eqnarray}
Here the order parameter carries momentum $\mathbf{q}$, with the
self-consistency condition
$\Delta_\mathbf{q} = g\sum_\mathbf{k} \langle
c_{\mathbf{k}\downarrow}c_{-\mathbf{k}+\mathbf{q}\uparrow}\rangle$. As
usual, the constant term related to the condensation energy has been
dropped from the reduced Hamiltonian Eq.~(\ref{eq:19}). Setting
$\mathbf{q}=0$ will reduce to the polarized BCS case \cite{Sarma63}.

As mentioned in Ref.~\cite{Frank}, a very important, and still open
issue, is the precise nature of the ground state in the regime
$h_\text{c} < h < h_\text{s}$, where $h_\text{c}=\Delta_0/\sqrt{2}$ sets the CC
transition. Let us now investigate the possible FFLO phase that may
arise in the intermediate region. Suppose we are in the normal FP
phase at some $h > h_\text{s}$, and the ``field'' $h$ is decreased until it
enters the PP phase. In order to have a qualitative and quantitative
description of this picture, for small $|\Delta_\mathbf{q}|$ one may
expand the action in fluctuations $|\Delta_{\vec{q}}|$ {\it a la}
Landau, since the transition from the FP to the normal-mixed phase is
continuous~\cite{Marenko,Simons,PRBMucio}.
We then expand the action up to the second order in the order parameter
$|\Delta_{q}|$~\cite{Marenko,Simons,PRBMucio}, and obtain
\begin{equation}
  S_{\text{eff}}=\sum_{\vec{q},\Omega}\alpha({|\vec{q}|}, \Omega) |\Delta_{\vec{q}}|^{2}+
  \mathcal{O}\left(|\Delta_{\vec{q}}|^4\right)\,,
\end{equation}
where
$\alpha(|{\vec{q}}|,\Omega)={1}/{g}-\chi({\vec{q}}, \Omega)$,
with $({\vec{q}},\Omega)$ being the four momentum of pairs, and
$\chi({\vec{q}}, \Omega)$ is the bare pair susceptibility without feedback
effect,
\begin{eqnarray}
\label{chi1-1}
\chi({\vec{q}}, \Omega)=
\sum_{\vec{k}} \frac{1-f(\xi_{\vec{k}-\vec{q}/2,\uparrow})-f(\xi_{\vec{k}+\vec{q}/2,\downarrow})}{\xi_{\vec{k}-\vec{q}/2,\uparrow}+\xi_{\vec{k}+\vec{q}/2,\downarrow}-\Omega},
\end{eqnarray}
where
$f(x)=1/(e^{\beta x}+1)$ is the Fermi
distribution function with $\beta \equiv 1/k_\text{B}T$.
Here $\chi({\vec{q}}, \Omega)$ can be obtained from analytical
continuation of the thermal pair susceptibility,
$\chi(Q)\equiv\chi(\mathbf{q}, i\Omega_l) $,
\begin{equation}
  \chi(Q) = \frac{1}{\beta}\sum_{\mathbf{k},i\omega_l} \mathcal{G}_0^\uparrow(K)\mathcal{G}_0^\downarrow(Q-K)\,,
  \label{eq:chi}
\end{equation}
where
$\mathcal{G}_0^\sigma(K) = (i\omega_l -\xi_{\mathbf{k}\sigma})^{-1}$
is the bare thermal Green's function. Here the four-vector
$K \equiv (\mathbf{k},i\omega_l)$ and
$Q \equiv (\mathbf{q},i\Omega_l)$, where $\omega_{l}=(2l+1)\pi/\beta$
and $\Omega_{l}=2l\pi/\beta$ are the fermionic and bosonic Matsubara
frequencies, respectively.

Apparently, for an
isotropic system, $\chi({\vec{q}}, \Omega)$ does not depend on the
direction of $\mathbf{q}$.  Evaluation of the equation above is
straightforward. At zero temperature we find that
$\chi({\vec{q}}, \Omega)$ at zero frequency is given by
%
\begin{eqnarray}
\label{chi1-2}
\chi({\vec{q}},0) &=&
                      N(0) \left\{ 1 + \ln \left(\frac{2\omega_\text{c}}{2 h} \right)\right.\nonumber\\
  &&{}- \left.\frac{1}{2}\left[ \ln \left| 1- {\bar q}^2 \right| + \frac{1}{ \bar q} \ln \left| \frac{1+ {\bar q}}{1- {\bar q}} \right| \right] \right\},
\end{eqnarray}
where $\omega_\text{c}$ is an energy cutoff, $N(0)=\dfrac{m k_\text{F}^{}}{2 \pi^2}$ is the density of states at the Fermi level for a single spin component, $\bar q \equiv \dfrac{v_\text{F}^{} q}{2h}$ is the dimensionless ``measure'' of the pair momentum, with $q \equiv |\vec{q}|$ and $v_\text{F}^{}$ is the Fermi velocity. 

\begin{figure}
\centerline{
\includegraphics[clip,width=3.4in]{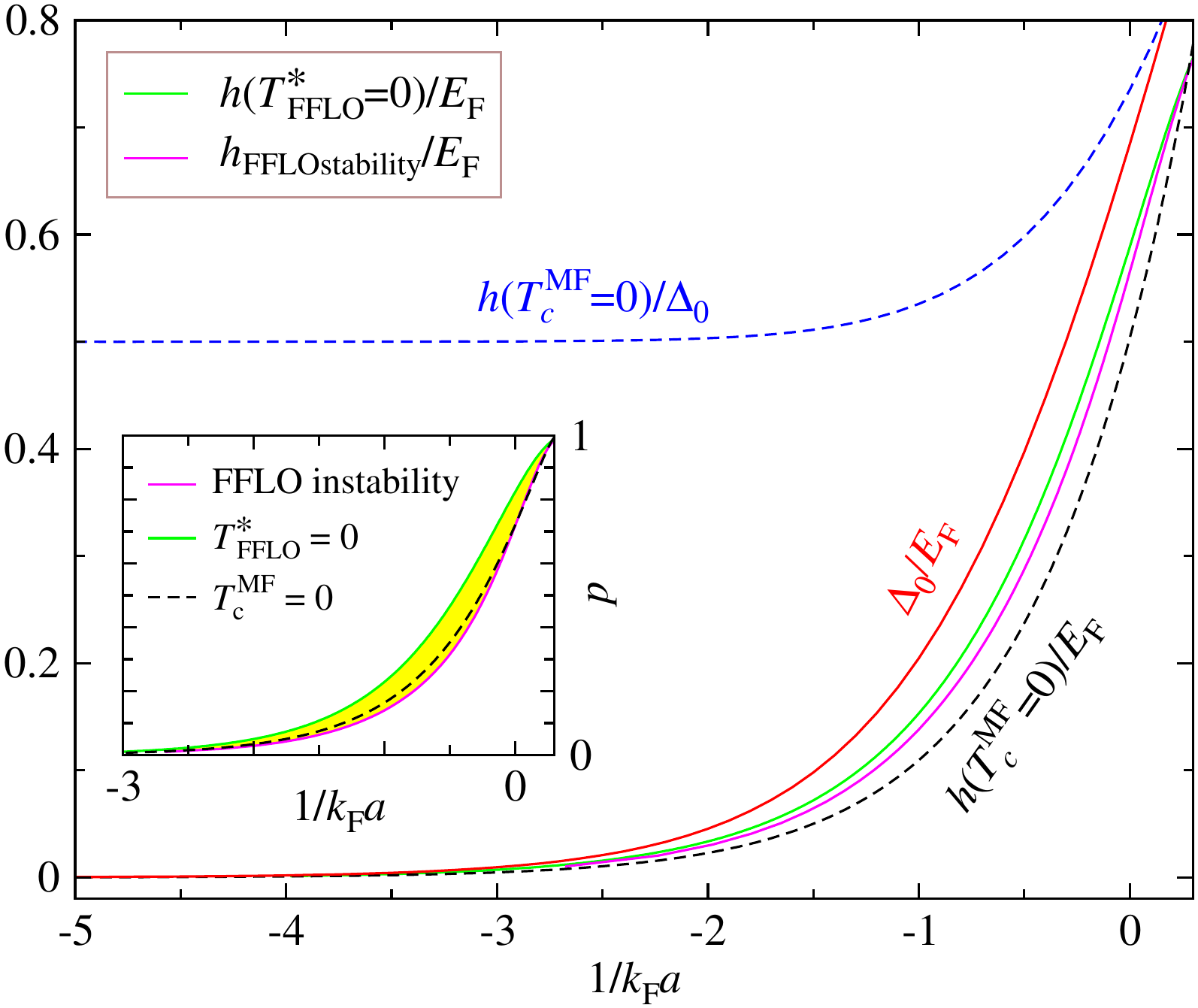}}
\caption[]{Gap $\Delta_0$ (red, as labeled) and $h$ along different
  lines, all in units of $E_\text{F}$, as well as their ratio
  $h(T_\text{c}^\text{MF}=0)/\Delta_0$ (blue dashed line), as a function of
  $1/k_\text{F}a$ for a 3D homogeneous Fermi gas. The calculation was done at
  the mean-field level without the particle-hole channel, which does
  not affect the ratio. Shown in the inset is the phase diagram of the
  stable FFLO phase. The upper boundary (green curve) is given by
  $T^*_\text{FFLO} =T_\text{c}^\text{FFLO}(\Delta=0)$, and the lower boundary
  (magenta curve) is given by the instability condition against phase
  separation. The Sarma $T_\text{c}^\text{MF}=0$ line (black dashed) lies
  within the FFLO phase. The $h$ value
  along these lines are shown in the main figure with the same color
  coding and line shapes.}
\label{fig:TestCC}
\end{figure}

The Thouless criterion for pairing instability, which corresponds to the divergence of the $T$ matrix $t(\mathbf{q},\Omega)$,
\begin{eqnarray}
\label{Thouless-criterion}
t^{-1}(\mathbf{q},0)=-\frac{1}{g}+\chi({\vec{q}},0)=0,
\end{eqnarray}
 yields,
\begin{eqnarray}
\label{Thouless}
\frac{h}{\Delta_0} = \frac{e}{2|1+\bar q|} \left| \frac{1 + \bar q}{1 - \bar q} \right|^{\frac{\bar q-1}{2\bar q}} = \frac{1}{2} + \frac{\bar{q}^2}{12} + \mathcal{O}(\bar{q}^3),
\end{eqnarray}
where $\Delta_0=2 \omega_\text{c} {\rm exp}(-1/N(0)g)$ is the zero
temperature BCS gap.  For a contact potential in 3D, which is relevant
for Fermi gases, one needs to replace the interaction strength $g$
with the dimensionless parameter $1/k_\text{F}a$ via the
Lippmann-Schwinger equation,
${m}/{4\pi a} = -1/g + \sum_\mathbf{k} 1/2\epsilon_\mathbf{k}$, where
$a$ is the two-body scattering length. Then the zero temperature gap
is given by $\Delta_0= \frac{8}{e^2} e^{\pi/2k_\text{F}a}$.  Note that
the exact expression for $\Delta_0$ is not crucial here, although
these specific expressions are appropriate only for the weak coupling
BCS regime. At the same time, it has been known that a possible FFLO
phase mainly exists on the BCS side of unitarity.  As shown in the
inset of Fig.~\ref{fig:TestCC}, at high imbalances, it  extends
slightly into the BEC side \cite{FFLO_MF_us}.

We determine the critical reduced momentum $\bar q_\text{c}$ by imposing an
extremal condition on the pair susceptibility.  Thus, extremizing
$\bar{\chi}(\bar{q})\equiv \chi({\vec{q}},0)$ with respect to $\bar q$ yields
\begin{eqnarray}
\label{qmin}
2 \bar q_\text{c} = \ln \left| \frac{1+ \bar q_\text{c}}{1- \bar q_\text{c}} \right|.
\end{eqnarray}
A numerical solution of the equation above gives $\bar q_\text{c} = 0$, and
$\bar q_\text{c} \simeq 1.2$. However, the locus of continuous transitions
may be determined from the value of $\bar q_\text{c}$ at which
$\alpha(\bar q_\text{c})$ is both minimized and passes through zero, and this
happens only for $\bar q_\text{c} \simeq 1.2$~\cite{Simons} or, equivalently,
at a wave-vector $q_\text{c} \simeq 2.4 h/v_\text{F}$.

The $\bar q = \bar q_\text{c}$ limit of $\bar{\chi}(\bar q_\text{c})$ gives
$N(0) \left[1 - 0.59 + \ln \left(\frac{2 \omega_\text{c}}{2 h} \right)
\right]$, so that $\alpha(q= q_\text{c}, 0)=0$ is
$-1/g+\bar{\chi}(\bar q = \bar q_\text{c})=-1/g + N(0) \left[0.41 + \ln
  \left(\frac{2\omega_\text{c}}{2 h} \right) \right]=0$, which leads to
$h_\text{s} \simeq 0.75 \Delta_0$, for the location of the FFLO transition,
agreeing with the findings of Shimahara~\cite{Shimahara}, Burkhardt and Rainer~\cite{Rainer}, and
Combescot and Mora~\cite{Combescot}.

The critical $h_\text{s}$ in turn yields the magnitude of the wave-vector
$q_\text{c} \simeq 1.8 \Delta_0/v_\text{F}$. These results also agree with the ones
obtained in Ref.~\cite{Shimahara} by a variational approach for a
three-dimensional FF superconductor with a spherical symmetric Fermi
surface. The FFLO window is then $h_\text{c} < h < h_\text{s}$, where the phase
transition at $h_\text{c} = h_\text{CC} \approx 0.71 \Delta_0$, is of first
order, and that at $h_\text{s} \simeq 0.75 \Delta_0$ is of second order. The
same results and conclusions are obtained when the calculations are
performed with the interaction $g$ replaced by $1/k_\text{F}a$ which is
appropriate for a short range interaction~\cite{Hu}, as they should.

It should be mentioned that the original Clogston derivation equates
the free energy of the superfluid state at zero-field (i.e., $h=0$)
with that of a polarized normal state at the threshold $h_\text{c}$,
both at $T=0$. This approach is expected to be valid for the small
$\Delta_0$ case in the perturbative sense. However, we argue that the
balanced and the imbalanced cases are really distinct and cannot
connect to each other continuously. This can be told from the fact
that in the BCS regime, an arbitrarily small but nonzero population
imbalance is sufficient to destroy superfluidity at precisely $T=0$ in
the 3D homogeneous case (when stability is taken into account)
\cite{Chien06}. For a finite $\Delta_0$, the ``magnetic field'' $h$
would have to jump from 0 of the balanced case to a value comparable
to $\Delta_0$, implying that $h$ should not be treated
perturbatively. Furthermore, there is no guarantee that the normal
state in Clogston's approach is a solution of the BCS gap equation in
the zero gap limit. To check the CC limit, we calculate for a 3D
homogeneous Fermi gas the gap $\Delta_0$ in the balanced case at zero
$T$ and the field $h$ in the imbalanced case when the mean-field
$T_\text{c}$ (also referred to as $T^*$) approaches 0, both as a
function of pairing strength. The result is shown in
Fig.~\ref{fig:TestCC}, where we plot $\Delta_0$ for $h=0$ (red curve)
and $ h$ along the $T_\text{c}^\text{MF}=0$ curve (black dashed), as
well as their ratio $h(T_\text{c}^\text{MF}=0)/\Delta_0$ (blue
dashed), as a function of $1/k_\text{F}a$.  This $h$ should be taken
as $h_\text{c}$ since it is the boundary between a normal phase and
polarized superfluid (also referred to as Sarma phase \cite{Sarma63})
with $q=0$ at $T=0$. The $T_\text{c}^\text{MF}=0$ curve for the Sarma
phase in the $p$ -- $1/k_\text{F}a$ plane can be easily obtained from
the BCS-like mean-field $T_\text{c}$ equation with $\mathbf{q}=0$ and
$T_\text{c}=0$, along with the fermion number constraints
\cite{He2007}. The figure indicates that the exact mean-field solution
yields $h_\text{c}/\Delta_0 = 0.5 $ in the BCS regime, substantially
different from $1/\sqrt{2}$ given by CC, and this ratio increases to
about 0.733 at unitarity. This result suggests that exact calculation
is needed in order to obtain quantitatively accurate value for
$h_\text{c}$.  It corresponds to the $q=0$ limit of
Eq.~(\ref{Thouless}) and is not stable. The difference between this
result and that of CC can likely be attributed to the possibility that
the CC normal state does not satisfy the Thouless criterion while the
present case does.

In the inset of Fig.~\ref{fig:TestCC}, we show the stable FFLO phase
at the mean-field level, as the yellow shaded region. The upper
boundary (green curve) is given by the zero gap solution,
$T^*_\text{FFLO}=0$, with a finite $q$ vector, which separates the
FFLO phase from the normal Fermi gases. The lower phase boundary
(magenta curve) is given by the instability condition of the FFLO
phase against phase separation. Both boundary lines were taken from
Ref.~\cite{He2007}. Next to but on the lower right side of this
boundary are phase separated states. It is clear that the Sarma
mean-field $T_\text{c}$ curve line (black dashed) lies completely within the
stable FFLO phase, in agreement with the fact that the Sarma states
along this curve are unstable against FFLO. We plot $h$ along these
two boundaries (green and magenta solid curves) in the main
figure. Interestingly, it turns out that, in the BCS limit, the ratio
$h/\Delta_0$ along the lower boundary is close to $1/\sqrt{2}$, in
agreement with Ref.~\cite{Casalbuoni}. Meanwhile, the ratio along the
upper boundary is close to 0.75. This leaves us with roughly the same
FFLO window of $0.71 < h/\Delta_0 < 0.75$ in the absence of the
induced interactions.
\vspace*{2ex}

\begin{figure}
\begin{centering}
\includegraphics[clip,width=6cm]{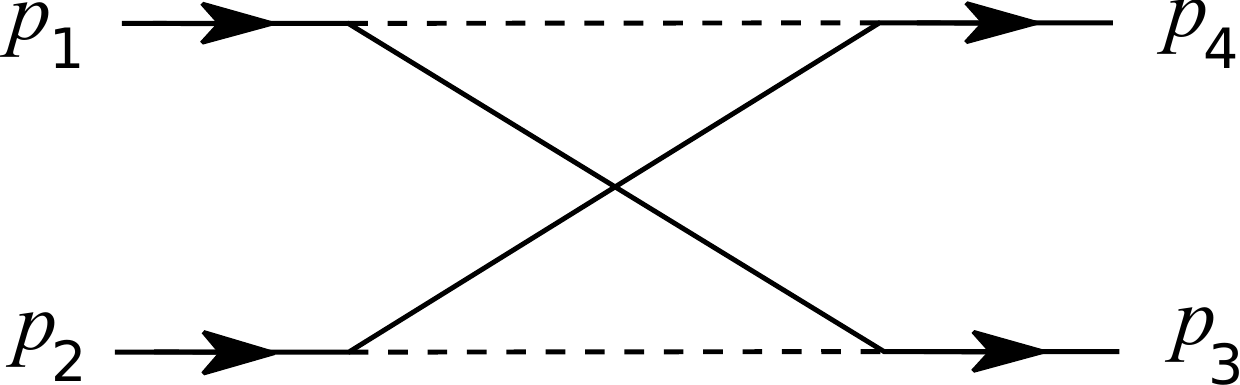}
\par\end{centering}
\caption[]{\label{LOD} The lowest-order diagram representing the
  induced interaction $U_{\text{ind}}( p_1, p_4)$. The solid and
  dashed lines represent fermion propagators and the interaction $g$
  between fermions, respectively.}
\end{figure}

\section{Effects of the Induced Interaction on the FFLO window}
\label{ind}

The induced interaction was obtained originally by GMB in the BCS
limit by the second-order perturbation \cite{GMB61}. For a scattering
process with $p_1+p_2\rightarrow p_3+p_4$, the induced interaction for
the diagram in Fig.~\ref{LOD} is expressed as
\begin{equation}
U_{\text{ind}}( p_1, p_4)= -g^2 \chi_\text{ph}(p_1-p_4),
\end{equation}
where $p_{i}=(\vec{k}_{i}, \omega_{l_i})$ is a four vector.
Including the induced interaction, the effective pairing interaction
between atoms with different spins is given by
\begin{eqnarray}
\label{induce}
U_\text{eff}( p_1, p_4)\equiv U_\text{eff}&=&
-g +U_{\text{ind}}( p_1, p_4)\\
\nonumber
&=& -g -g^2 \chi_\text{ph}(p_1-p_4).
\end{eqnarray}
The polarization function $\chi_\text{ph}(p')$ is given by
\begin{eqnarray}
  \chi_\text{ph}(p')&=& \sum_p
                   \mathcal{G}_{0}^b(p)\mathcal{G}_{0}^a(p+p')\nonumber\\
               &=&\int {{\rm{d}}^3k \over (2\pi)^3} \frac{f_{{\vec k}}^b-f_{{\vec k}+{\vec q}}^a} {i\Omega_l+\xi_{{\vec k},a}-\xi_{{\vec k}+{\vec q},b}},
\label{chi0}
\end{eqnarray}
where $f_{{\vec k}}^\sigma\equiv f(\xi_{\vec{k},\sigma})$, and
$p'=(\vec{q}, \Omega_{l})$.  This means that $U_\text{eff}$ is a
function of momentum and frequency. The static polarization function
is then,
\begin{widetext}
\begin{eqnarray}
\label{chiph-5}
\chi_\text{ph}(q,h)
&=& - \frac{2m}{(2 \pi)^2} \int \frac{\mathrm{d} k\, k^2}{2qk}  \left[  f_{k}^{\downarrow} \ln \left( \frac{q^2 -4mh+2kq }{q^2 -4mh-2kq} \right)+  f_{k}^{\uparrow} \ln \left( \frac{q^2 +4mh+2kq }{q^2 +4mh-2kq} \right)   \right],
\end{eqnarray}
where $q \equiv |\vec q|$. The above expression is usually computed in the zero temperature limit, with $ f_{k}^{\downarrow,\uparrow} \to  \Theta(k_\text{F}^{\downarrow,\uparrow}-k)$, where $\Theta(x)$ is the step function, such that the induced correction to the coupling $g$ is a (temperature independent) constant.
%
\begin{eqnarray}
\label{chiph-6}
\chi_\text{ph}(q,h) &\equiv& \chi^{\downarrow}(q,h) + \chi^{\uparrow}(q,h)\\
\nonumber
&=& - \frac{m}{(2 \pi)^2 q}  \int_0^{k_\text{F}^{\downarrow}} \mathrm{d} k\, k    \ln \left( \frac{q^2 -4mh+2kq }{q^2 -4mh-2kq} \right)
{}- \frac{m}{(2 \pi)^2 q}  \int_0^{k_\text{F}^{\uparrow}} \mathrm{d} k\, k  \ln \left( \frac{q^2 +4mh+2kq }{q^2 +4mh-2kq} \right).
\end{eqnarray}
Equation~(\ref{chiph-6}) shows that the static polarization function in the case of a spin imbalanced Fermi gas separates into contributions from the spin-down and the spin-up like susceptibilities. The integration in $k$ gives
\begin{eqnarray}
\label{chiph-7}
\chi_\text{ph}^{\downarrow}(q,h) &=& -\frac{m}{ 8 \pi^2 q} \left\{      \left[ {k_\text{F}^{\downarrow}}^2-\left( \frac{q^2-4mh}{2q} \right)^2 \right ]\ln \left| \frac{q^2 -4mh+2q k_\text{F}^{\downarrow} }{q^2 -4mh-2q k_\text{F}^{\downarrow}} \right| + k_\text{F}^{\downarrow}\left( \frac{q^2-4mh}{q} \right)    \right\},\\
\label{chiph-8}
\chi_\text{ph}^{\uparrow}(q,h) &=& -\frac{m}{ 8 \pi^2 q} \left\{    \left[ {k_\text{F}^{\uparrow}}^2-\left( \frac{q^2+4mh}{2q} \right)^2 \right ] \ln \left| \frac{q^2 +4mh+2q k_\text{F}^{\uparrow} }{q^2 +4mh-2q k_\text{F}^{\uparrow}} \right| + k_\text{F}^{\uparrow}\left( \frac{q^2+4mh}{q} \right)    \right\}.
\end{eqnarray}
%
The equations above can be put in a more convenient form, $\tilde{\chi}_\text{ph}^\sigma(x,y)\equiv \chi_\text{ph}^\sigma(q,h)$, where
%
\begin{eqnarray}
\label{chiph7-2}
\tilde{\chi}_\text{ph}^{\downarrow}(x,y) &=& -\frac{N(0)}{4} \left\{\sqrt{1-y}\left(1 - \frac{y}{2x^2} \right)  -  \frac{1}{2x} \left[ 1 - y - x^2 \left( 1- \frac{y}{2x^2} \right)^2 \right ]  \ln \left| \frac{\sqrt{1-y} + \frac{y}{2x} - x }{\sqrt{1-y} - \frac{y}{2x} + x} \right|   \right\} \\
\nonumber
&=& - N(0) L^{\downarrow}(x,y),
\end{eqnarray}
and
\begin{eqnarray}
\label{chiph8-2}
\tilde{\chi}_\text{ph}^{\uparrow}(x,y) &=& -\frac{N(0)}{4} \left\{  \sqrt{1+y}\left(1 + \frac{y}{2x^2} \right)    -  \frac{1}{2x} \left[ 1 + y - x^2 \left( 1+ \frac{y}{2x^2} \right)^2 \right ] \ln \left| \frac{\sqrt{1+ y} - \frac{y}{2x} - x }{\sqrt{1+ y} + \frac{y}{2x} + x} \right|  \right\} \\
\nonumber
&=& - N(0) L^{\uparrow}(x,y),
\end{eqnarray}
\end{widetext}
where
$x\equiv \frac{q}{2 k_\text{F}}$, and $y\equiv \frac{h}{\mu}$. This allows us
to write the polarization function of an imbalanced Fermi gas as
\begin{eqnarray}
\label{generalizedL}
\tilde{\chi}_\text{ph}(x,y) = - N(0) L(x,y),
\end{eqnarray}
where $L(x,y) \equiv L^{\downarrow}(x,y) + L^{\uparrow}(x,y)$ is the generalized Lindhard function.

Notice that in the $y \to 0$ limit, $k_\text{F}^{\downarrow}=k_\text{F}^{\uparrow}=k_\text{F}$, such that $\tilde{\chi}_\text{ph}^{\downarrow}(x,0) = \tilde{\chi}_\text{ph}^{\uparrow}(x,0) \equiv \tilde{\chi}_\text{ph}(x)/2$ and we obtain the well-known (balanced) result
\begin{eqnarray}
\nonumber
\tilde{\chi}_\text{ph}(x) &=&- \frac{m}{ 4 \pi^2 q} \left[  k_\text{F}  q -   \left( k_\text{F}^2- \frac{q^2}{4}  \right ) \ln \left| \frac{q^2 - 2q k_\text{F} }{q^2 + 2q k_\text{F}} \right|   \right]\\
&=& -N(0)  L(x),
\label{chiph-9}
\end{eqnarray}
where $ L(x)\equiv L(x,0)$ is the standard Lindhard function,
\begin{eqnarray}
\label{chiph9-2}
 L(x)= \frac{1}{2} -\frac{1}{4x}(1-x^2) \ln \left|\frac{1-x}{1+x} \right|.
\end{eqnarray}

In the scattering process the conservation of total momentum implies
that $\vec{k}_1+\vec{k}_2=\vec{k}_3+\vec{k}_4$, with
$\vec{k}_1=- \vec{k}_2$ and $\vec{k}_3=-\vec{k}_4$. The momentum $q$
is equal to the magnitude of $\vec{k}_1+\vec{k}_3$, so that
$q=\sqrt{(\vec{k}_1+\vec{k}_3).(\vec{k}_1+\vec{k}_3)} =
\sqrt{\vec{k}_1^2+\vec{k}_3^2+
  2\vec{k}_1. \vec{k}_3}=\sqrt{\vec{k}_1^2+\vec{k}_3^2+ 2|\vec{k}_1||
  \vec{k}_3|\cos \phi}$, where $\phi$ is the angle between $\vec{k}_1$
and $\vec{k}_3$. Since both particles are at the Fermi surface,
$|\vec{k}_1|=|\vec{k}_3|=k_\text{F}=\sqrt{2m\mu}$, thus,
$q=k_\text{F}\sqrt{2(1+\cos \phi)}$, and consequently
$x=\sqrt{2(1+\cos \phi)}/2$, which sets $0 \leq x \leq1$.

The {\it s}-wave part of the effective interaction is approximated by
averaging the polarization function $\chi_\text{ph}(q)$ over the Fermi
sphere, which means an average of the angle
$\phi$~\cite{Pethick00,Qijin,Petrov,Baranov2,Yu09,Torma,Yu10},
\begin{eqnarray}
\langle\tilde{\chi}_\text{ph}(x,y)\rangle &=& \frac{1}{2} \int_{-1}^{1} \mathrm{d} \cos\phi ~\tilde{\chi}_\text{ph}(x,y)\nonumber\\
&=& - N(0) \left[  \frac{1}{2} \int_{-1}^{1} \mathrm{d} \cos\phi ~ L(x,y)\right] \nonumber\\
&\equiv& - N(0) \bar L(y),
\end{eqnarray}
where we have made use of Eq.~(\ref{generalizedL}). The quantity
$\bar L$ characterizes the magnitude of GMB corrections in the
presence of population imbalance. Shown in the inset of
Fig.~\ref{h-over-Delta} is the behavior of $\bar L(y)$ as a function
of imbalance $y$. In the $y\rightarrow 0$ limit, we have precisely
$\bar L(0) =(1+2\ln 2)/3 = 0.795431454 $, as given in
Ref.~\cite{Qijin} and other papers \cite{Yu10} for the balanced
case. As $y$ increases from 0 to 1, $\bar L(y)$ decreases to 0.69,
indicating that the particle-hole fluctuation effect becomes weaker
due to the Fermi surface mismatch caused by population imbalance. This
result is identical to that of Ref.~\cite{Yu10}.

Taking into account the GMB correction, the divergence of the $T$ matrix in Eq.~(\ref{Thouless-criterion}) is now given by
\begin{equation}
  t^{-1}(\mathbf{q},0)=\left(-\frac{1}{g} + \langle\tilde{\chi}_\text{ph}(x,y)\rangle \right)+\chi({\vec{q}},0)=0,
  \label{eq:22}
\end{equation}
which can be obtained by replacing $g$ in
Eq.~(\ref{Thouless-criterion}) with $U_\text{eff}$, as given in
Eq.~(\ref{induce}). This expression has been shown to be correct when the more complicate $T$ matrix in the particle-hole channel is included self-consistently \cite{Qijin}. This yields a GMB corrected solution $\Delta_0^\text{GMB}$ satisfying $\dfrac{h(y)}{\Delta_0^\text{GMB}}=\left(\dfrac{h(y)}{\Delta_0}\right)^\text{MF}$ and $\Delta^\text{GMB}(y) = \Delta(y) e^{-\bar{L}(y)}$, with $\Delta_0 \equiv \Delta(y=0,T=0)$ and $\Delta_0^\text{GMB} = \Delta^\text{GMB}(y=0,T=0)$. This amounts to
\begin{equation}
\label{hcrit}
\frac{h_\text{s}}{\Delta_0}
= \left( \frac{h_\text{s}}{\Delta_0} \right)^\text{MF} e^{-\bar L(y_\text{s})},
\end{equation}
where $y_\text{s} = h_\text{s}/\mu$, and
$\left( \frac{h_\text{s}}{\Delta_0} \right)^\text{MF} \simeq 0.75$ is the MF
result without the GMB corrections. 

\begin{figure}
\centerline{\includegraphics[clip,width=3.4in]{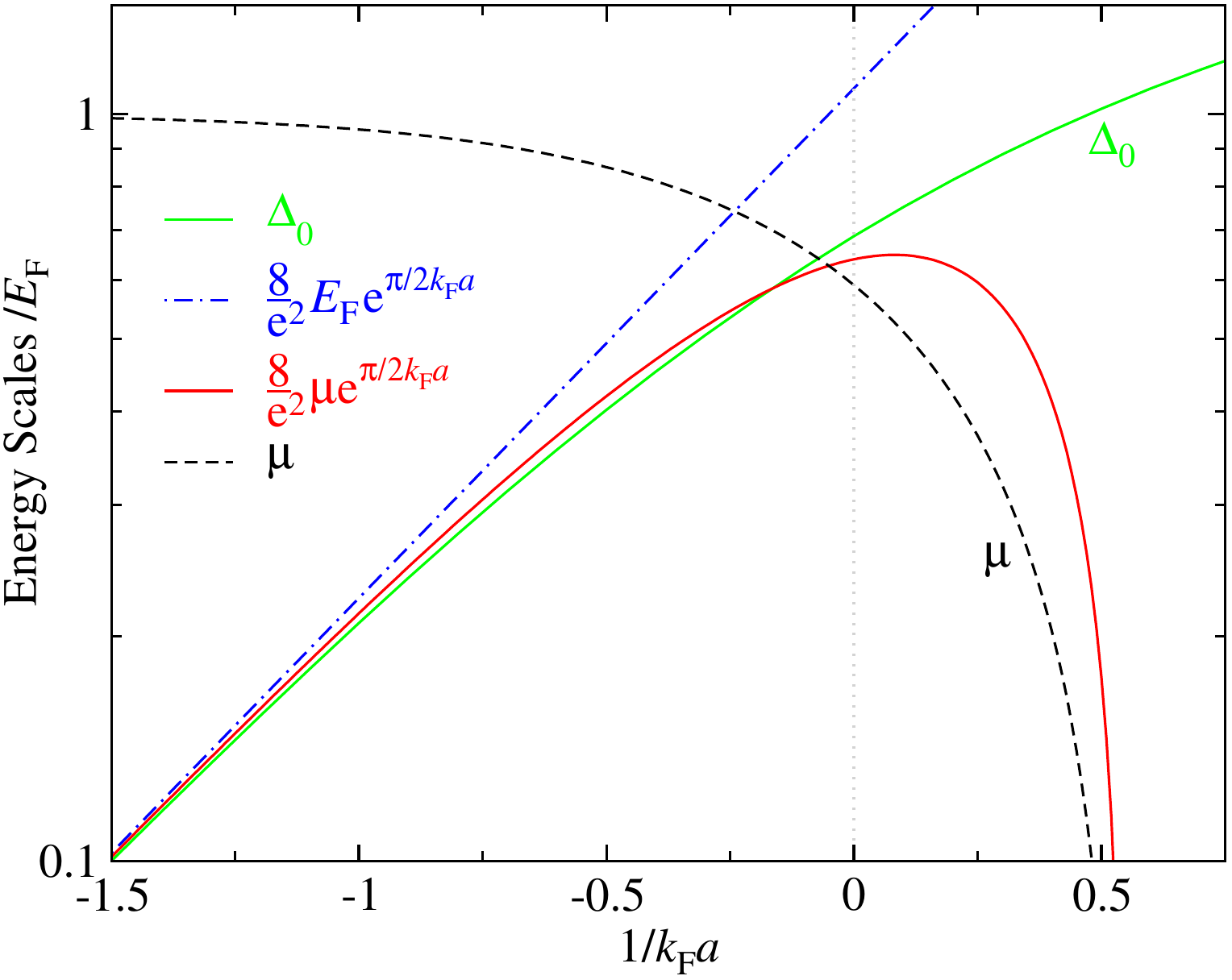}}
\caption{Behavior of self-consistently calculated zero temperature
  $\mu$ (black dashed) and $\Delta_0$ (green solid), as well as the
  weak coupling BCS expression (blue dot-dashed) and our corrected
  approximation (red solid line) for $\Delta_0$, in units of $E_F$, as
  a function of $1/k_Fa$ for a 3D homogeneous Fermi gas.}
\label{fig:mu}
\end{figure}

It is well known that the zero temperature BCS pairing gap (at $y=0$) is modified due to the particle-hole channel effect (or GMB correction) as~\cite{Pethick00,Qijin}
\begin{eqnarray}
\Delta_0^\text{GMB} &=& \frac{\Delta_0}{(4e)^{1/3}} = \frac{8}{e^2} \frac{1}{(4e)^{1/3}} \mu e^{-\pi/2k_\text{F}|a|}\nonumber\\
&=& \left(\frac{2}{e}\right)^{7/3} \mu e^{-\pi/2k_\text{F}|a|}.
\label{GMBgap}
\end{eqnarray}
Note here that in the expression for $\Delta_0$, the chemical
potential $\mu$ plays the role of $E_\text{F}$. This can be readily
obtained following the standard derivation in the BCS framework, but
allowing the Fermi level to evolve continuously as one does for the
BCS-BEC crossover \cite{Leggett1980}. This automatically corrects the
moving density of states as the Fermi level changes, even though the
approximation becomes less accurate by replacing the full momentum
space integral by an energy integral with the density of states fixed
at the Fermi level.
Shown in Fig.~\ref{fig:mu} is a comparison between the calculated
$\Delta_0$ and different analytical approximations. The blue
dot-dashed line is the expression in the weak coupling limit, with
$\mu$ pinned at $E_F$. Our corrected expression is shown as the red
solid curve. Both are to be compared with $\Delta_0$ (green solid
curve), which is calculated self-consistently in the context of
BCS-BEC crossover. It is evident that our corrected expression is
quantitatively good all the way from the BCS through the unitary
limit.

In order to find the appropriate $y$ for consistently evaluating
Eq.~(\ref{hcrit}), we take the expression for the GMB gap at
unitarity, $1/k_\text{F}a = 0$, and obtain
\begin{eqnarray}
\label{GMBgap2}
\Delta_0^\text{GMB} =  \left(\frac{2}{e}\right)^{7/3}\!\!\! \mu \simeq 0.49 \mu
\end{eqnarray}
for $y=0$.  Note that this is very close to the more complete solution
of $\Delta_0^\text{GMB}= 0.42E_\text{F} = 0.50 \mu$ with $\mu = 0.837E_\text{F}$,
calculated with the full particle-hole $T$ matrix included at the
$G_0G_0$ level \cite{Qijin}.
According to Eq.~(\ref{GMBgap}), when taking into account the GMB correction,
 the original  $h_\text{c}/\Delta_0=1/\sqrt{2}$ is transformed to
  $h_\text{c}^\text{GMB}/\Delta_0^\text{GMB}=1/\sqrt{2}$, which, with
  Eq.~(\ref{GMBgap2}), yields
  $h_\text{c}^\text{GMB}/\mu \equiv y_\text{c} = \left(\frac{2}{e}\right)^{7/3} \frac{1}{\sqrt{2}} = 0.3455$ at unitarity.

In Eq.~(\ref{hcrit}), we approximate $y_\text{s} \equiv h_\text{s}/\mu$ by $y_\text{c}$ in
$\bar{L}(y_\text{s})$, and obtain $\bar L = 0.7857$, and thus
\begin{equation}
\label{hcrit2}
\frac{h_\text{s}^\text{GMB}}{\Delta_0}  = 0.4558 \left( \frac{h_\text{s}}{\Delta_0}\right)^\text{MF} \simeq 0.342.
\end{equation}
Alternatively, one can take $\dfrac{h_\text{s}^\text{GMB}}{\Delta_0^\text{GMB}} = \left(\dfrac{h_\text{s}}{\Delta_0}\right)^\text{MF}$ and obtain immediately
\begin{equation}
\label{hcrit2a}
  \frac{h_\text{s}^\text{GMB}}{\Delta_0} = \frac{1}{(4e)^{1/3}} 0.75 = 0.339,
\end{equation}
which agrees with  Eq.~(\ref{hcrit2}).
On the other hand, with the shifted interaction strength, the CC limit
is modified to
\begin{eqnarray}
\label{hcrit3}
  \frac{h_\text{c}^\text{GMB}}{\Delta_0} &=& \frac{1}{(4e)^{1/3}} \frac{1}{\sqrt{2}} = 0.319.
\end{eqnarray}
Combining Eq.~(\ref{hcrit2}) and Eq.~(\ref{hcrit3}), we conclude that the screening of the medium (i.e., the induced interactions) has shrunk the FFLO window to
\begin{equation}
  0.32 \lesssim h/\Delta_0 \lesssim 0.34\,.
\label{ImprovedWindow}
\end{equation}

\section{Effects of induced interactions in the presence of impurities}
\label{Ap}

The above calculations in Section~\ref{ind} have been done assuming
the system is clean. However, this is not always true, especially for
a superconductor, for which impurities and dislocations are easy to
find. Impurities may cause a finite lifetime for the quasiparticles.
In quasi-one-dimensional organic superconductors, for instance, the
issue of lifetime effects arise from nonmagnetic impurities or
defects~\cite{Ardavan}. These impurities may add to the complexity of
the effect of particle-hole fluctuations, and thus deserve careful
inspection.

\begin{figure}[t]
\centerline{
\includegraphics[clip,width=3.3in]{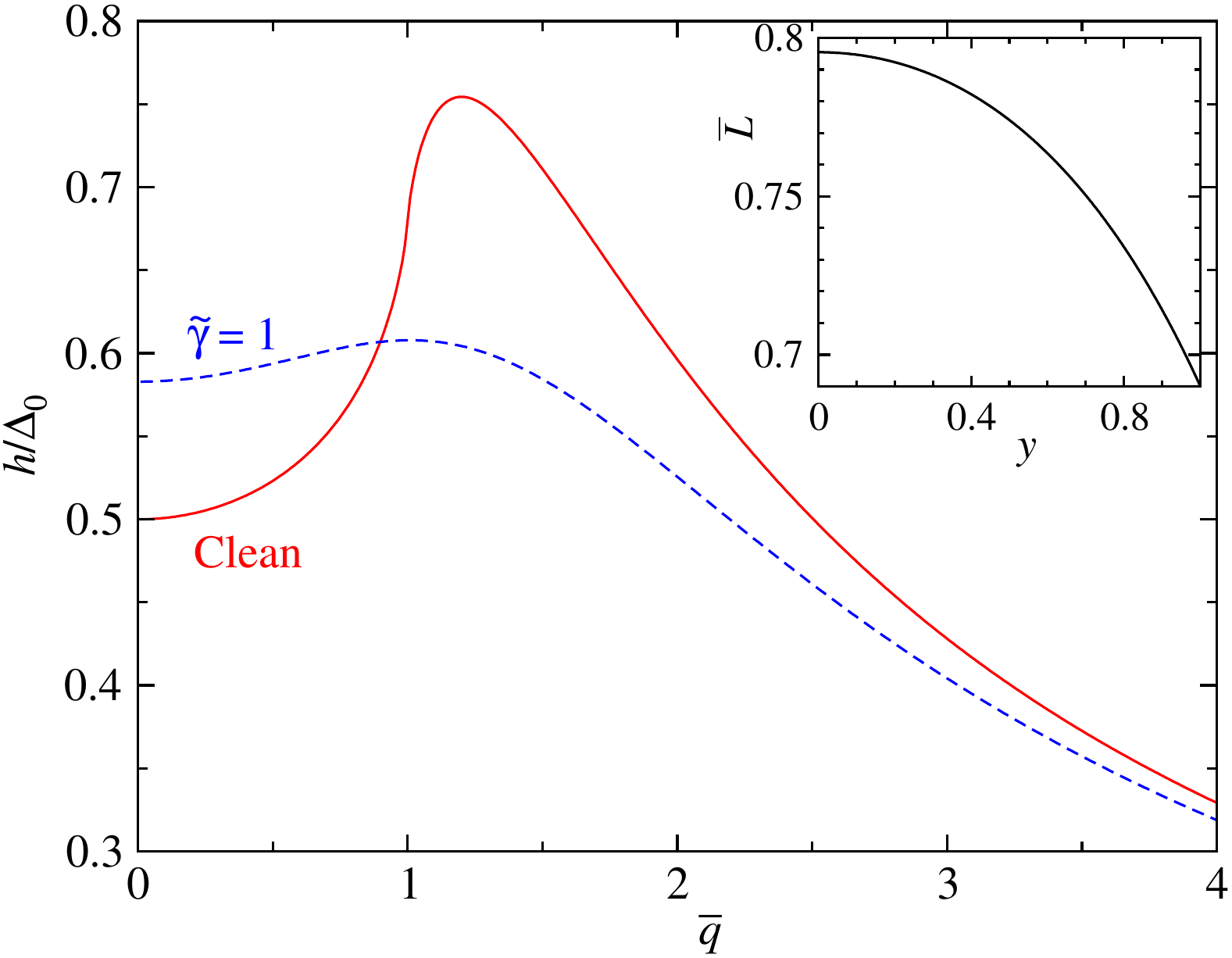}}
\caption[]{\label{h-over-Delta} The ratio $h/\Delta_0$ as a function of $q$.  The solid curve is for $\gamma=0$, and attains its maximum value $\simeq 0.75$, at $\bar q=\bar q_\text{c} \simeq 1.2$. The dashed curve is for $\tilde\gamma=1$ and shows a significant reduction of the maximum value of $h/\Delta_0$. Shown in the inset is the behavior of  $\bar L$ as a function of $y=h/\mu$.}
\end{figure}

In this section we show that, indeed, the FFLO window may be strongly
affected by impurity effects \cite{Agterberg_2001}. We shall only
consider nonmagnetic weak impurities in the Born limit
\cite{CheImpurity}, which mainly lead to a spectral broadening
$\gamma$ for the fermions \footnote{For nonmagnetic impurities in the
  Born limit, one has $\gamma = n_i u^2$, where $n_i$ is the impurity
  density, and $u$ is the impurity scattering strength. See, e.g.,
  Ref.~\cite{Abrikosov} for details.}. Then we rederive everything in
the presence of the spectral broadening. For the nonmagnetic
impurities which we consider, possible modification to the real part
of quasiparticle dispersion may be absorbed into the chemical
potential. Along with the simplification of the imaginary part by a
constant parameter $\gamma$, these nonmagnetic impurities satisfy the
Anderson's theorem in the BCS regime \cite{Anderson}\footnote{Such
  simplification is appropriate only for nonmagnetic impurities in the
  Born limit. One may see, e.g., Ref.~\cite{Chen-Schrieffer} for an
  example how Anderson's theorem may be broken when going beyond this
  simplified approximation.}. As the gap becomes large, the small gap
approximation assumed by Anderson's theorem is no longer
valid. However, a large $s$-wave gap itself is very robust against
weak impurities \cite{CheImpurity}.  In Fig.~\ref{h-over-Delta}, we
show the numerical solutions of $h/\Delta_0$ as a function of
$\bar{q}$ from Eqs.~(\ref{Thouless}) and (\ref{chi10-g10-3d}), for
both the clean limit i.e., $\gamma=0$ (red curve), and the dirty case
with $\tilde\gamma \equiv \gamma/2h=1$ (blue dashed line), where
$\gamma = \tau^{-1}$ is the inverse of the lifetime of a quasiparticle
in the normal phase. The clean case has a maximum value at
$\bar q=\bar q_\text{c} \simeq 1.2$, which gives
$h/\Delta_0 = h_\text{s}/\Delta_0 \simeq 0.75$, as obtained by the
analytical calculations in Section~\ref{N-M}. In the dirty case, the
maximum has shifted toward lower $\bar{q}$, and the maximum ratio
$h_\text{s}/\Delta_0$ has decreased significantly. This inevitably
narrows the FFLO window. When this ratio drops below $1/\sqrt{2}$, the
FFLO window will be gone and thus the FFLO phase will disappear at the
mean-field level.

Impurities in principle have an effect on the GMB correction, mainly
via changing the chemical potential $\mu$. However, we point out this
is only a minor secondary effect, since the change in $\mu$ due to
impurities is often very small, especially for Born impurities, which
cannot induce an impurity band outside the Fermi sphere
\cite{Chen-Schrieffer}. The GMB correction used in the present
calculation has been done at the lowest level approximation
\cite{Qijin}, without considering the gap effect at the Fermi
level. Therefore, we believe that at this level, one can safely
neglect the impurity effect on the GMB correction to the pairing
strength.
Therefore, we conclude that \emph{the effect of particle-hole
  fluctuations may be largely taken care of by assuming that it is
  encoded in an effective pairing strength}, \textit{a la}
Eq.~(\ref{eq:22}). One only needs to roughly rescale $h_s$ obtained in
the presence of impurities by same factors $(4e)^{-1/3}$ as in the
clean case. Hence, it does not have to appear explicitly in our
impurity derivations below.
It should be noted, however, that $h_\text{c}$ is unaffected by
impurities, for two reasons. As given by the Anderson's theorem,
$\Delta_0$ is unaffected by the weak nonmagnetic impurities. The
thermodynamics calculation of $h_\text{c}$, which involves the free
energy density of magnetic field, $H_0^2/8\pi$, as given by Clogston
\cite{Clogston}, is insensitive to impurities. Therefore, $h_\text{c}$
is only subject to the GMB correction.

Considering the finite lifetime of the quasi-particle states in the
momentum representation, the pair susceptibility $\chi$ is found, via
Eqs.~(\ref{eq:chi}) and (\ref{eq:22}), by the standard method of
including a finite imaginary part to the Green's
function~\cite{Leung,Abrikosov},
$\mathcal{G}_0^\sigma(K) =
1/(i\omega_l-\xi_{\vec{k}\sigma}+i\gamma\,\mathrm{sgn}(\xi_{\vec{k}\sigma}))$.
After somewhat lengthy but straightforward derivations (as shown in
Appendix \ref{ApA}), the real part of the particle-particle dynamic
pair susceptibility can be written as,
\begin{widetext}
\begin{eqnarray}
\label{chi10-g8-3d}
\mathrm{Re} \bar{\chi}({\bar{q}}) &=&
 N(0) \left[ 1 + \ln \left(\frac{2\omega_\text{c}}{2 h} \right) + \frac{1}{4 \bar q} \left( 1- {\bar q} \right)  \ln \left( (1- {\bar q})^2+\tilde\gamma^2 \right) - \frac{1}{4 \bar q} \left( 1+ {\bar q} \right)  \ln \left( (1+ {\bar q})^2+\tilde\gamma^2 \right) \right],
\end{eqnarray}
\end{widetext}
which is the counterpart of Eq.~(\ref{chi1-2}).
This approach is formally close to that used to investigate the effect
of non-magnetic impurities in one dimensional imbalanced Fermi
gases~\cite{Continentino}, and in two~\cite{Mineev} and
three~\cite{Takada_1970} dimensional FFLO superconductors.

With $\text{Re} \bar\chi({\bar{q}})$ from Eq.~(\ref{chi10-g8-3d}) the divergence of the $T$ matrix now yields
\begin{eqnarray}
\label{chi10-g10-3d}
\frac{h}{\Delta_0} = \frac{e}{2} \left[ (1- \bar q)^2 +\tilde\gamma^2 \right]^{\mathlarger{\frac{1- \bar q}{4 \bar q}}} \left[ (1+ \bar q)^2 + \tilde\gamma^2\right]^{-\mathlarger{\frac{1+ \bar q}{4 \bar q}}}.\hspace*{1cm}
\end{eqnarray}
Notice that in the limit $\tilde\gamma \to 0$ in Eqs.~(\ref{chi10-g8-3d}) and~(\ref{chi10-g10-3d}), the standard results in Eqs.~(\ref{chi1-2}) and~(\ref{Thouless}) are recovered.

Instead of being a solution of Eq.~(\ref{qmin}), the critical reduced
momentum $\bar q_\text{c}$ is now given by the solution of
\begin{eqnarray}
\label{qmin2}
 2 \bar q_\text{c} \left[ \frac{(1+ \bar q_\text{c})^2 }{(1+ \bar q_\text{c})^2 + \tilde \gamma^2}\right.& +& \left.\frac{(1- \bar q_\text{c})^2}{(1- \bar q_\text{c})^2 + \tilde \gamma^2} \right] \\ \nonumber &&= 
 \ln \left[ \frac{(1+ \bar q_\text{c})^2 + \tilde \gamma^2}{(1- \bar q_\text{c})^2 + \tilde \gamma^2} \right].
\end{eqnarray}
For $\tilde \gamma=1$, for instance, the numerical solution of the
equation above yields $\bar q_\text{c} \simeq 1.01$, besides the
trivial solution $\bar q_\text{c} = 0$. This leads to
$h_\text{s} \simeq 0.61 \Delta_0$ for the location of the FFLO
transition, (which transforms to
$h_\text{s}^\text{GMB} \approx 0.28\Delta_0$), in agreement with
Fig.~\ref{h-over-Delta}. However, this value is beyond the critical
value $\tilde \gamma_\text{c}=0.3$, which gives
$h_\text{s}^\text{critical}/\Delta_0 \simeq 0.71 =
h_\text{c}/\Delta_0$, or
$h_\text{s}^\text{GMB,critical} = 0.32\Delta_0 =
h_\text{c}^\text{GMB}$, for closing the FFLO window. This means that
at this critical value of $\gamma$ the system undergoes a first-order
quantum phase transition from the BCS to the polarized normal
phase. Conversely, with infinite life time ($\gamma=0$) the FFLO
window remains open with the ``unperturbed'' limits
$h_\text{c}^\text{GMB} < h < h_\text{s}^\text{GMB}$, as given by
Eq.~(\ref{ImprovedWindow}). This nontrivial result comes from the fact
that $h_\text{c}$ and $ h_\text{s}$ respond to impurities differently.

\section{Further Discussions}

It should be noted that we have in fact considered only the FF
case. This is justified in that there is no self-consistent way to
calculate the LO and higher order crystalline LOFF phases when the
pairing gap is large.  In the original LO paper \cite{LO}, the LO
state order parameter is treated as a small perturbation to the
noninteracting fermion propagator so that in the evaluation of all the
diagrams,  the Green's function is treated at the
noninteracting level. Such a perturbative treatment necessarily breaks
down in the unitary regime, where the gap is large, comparable to the
Fermi energy. In addition, in the presence of two wavevectors
$\pm\mathbf{q}$, a simple diagrammatic analysis shows that it will
generate an infinite series of components of wavevector
$\pm n \mathbf{q}$ ($n = 0, 1, 2, ...$) in the order parameter
\cite{Datta_2019PRB}. Indeed, many later works on LO and higher order crystalline states treat the order parameters as an expansion parameter, in a Ginzburg-Landau type of formalism \cite{BuzdinPLA,Agterberg_2001,Adachi_2003PRB,Vorontsov_2008PRB}, and thus they are appropriate only in the weak pairing regime. Therefore, unlike the FF case, there is no
simple field-theoretical approach to the LO and higher order
crystalline phases beyond the perturbative mean-field treatment with a
truncation of the series of the wavevectors \cite{chen07prb}. This makes
it more difficult to include the GMB or particle-hole fluctuation effect
using field-theoretical techniques \cite{Yu09,Yu10,Qijin}.  It
remains a challenging issue for the future to treat self-consistently
the effect of particle-hole fluctuations on the LO and higher order
crystalline FFLO states beyond the mean-field level.

In the presence of inhomogeneity, the Bogoliubov -- de Gennes (BdG)
treatment is often used \cite{YangPRB2008}. It is particularly useful
for treating Fermi gases in a trap. However, it should be emphasized
that BdG is also a mean-field treatment, albeit in real space.  In the
presence of multiple wavevectors, the complexity of the generalized
formalism increases rapidly (see, e.g., \cite{Datta_2019PRB}). There
has been thus far no report in the literature of incorporating
particle-hole fluctuations in the BdG formalism.

It should be pointed out that, for population imbalanced Fermi gases
in a trap, the local population imbalance $p$ varies as a function of the radius. While
an inverted density distribution is possible \cite{Liao,wang13pra}, in
most cases, $p$ increases from zero (or nearly zero depending on the
temperature) at the trap center to unity at the trap edge. One may
think of the radius as an equivalent of the imbalance $p$ in the inset of
Fig.~\ref{fig:TestCC}. At low $T$, except for the BEC regime, one may
find that the FFLO solution exists at certain radius, or inside a
narrow shell near this radius, under the local density approximation
(LDA). The thickness of the shell, relative to the coherence length,
may have a strong influence as to whether a FFLO solution exists. In
such a case, BdG may have an advantage over LDA.

Our impurity treatment has been restricted to nonmagnetic impurities
in the Born limit, following the approach of Anderson \cite{Anderson}
and Abrikosov and Gor'kov \cite{AG-JETP,Abrikosov}. This assumes 
randomly distributed weak impurities, whose effect can mainly be
simplified as a finite life time effect in the quasiparticles. 
Recent works \cite{LiHan-NJP,CheImpurity} show that weak disorders do not
significantly affect $T_c$ of an $s$-wave BCS superfluid in accordance with Anderson's theorem \cite{Anderson}, and the
superfluid is more robust to the presence of disorder in the unitary
regime. There has been treatment beyond the Born limit, in the context
of $d$-wave high $T_c$ superconductors
\cite{Hirschfeld,Chen-Schrieffer,Zhu-RMP} and $s$-wave atomic Fermi
gases in the BCS-BEC crossover \cite{Strinati_2013PRB,CheImpurity}.  There are
of course also treatments of pair breaking, magnetic disorders or
impurities in superconductors
\cite{Fowler1970,Maki1972,Bickers1987,Vernier2011,Riegger2018}. It
would be interesting to investigate how an FFLO phase responds to
magnetic impurities.  Apparently, the impurity averaging technique has
been widely applied to the impurity treatment for the FFLO states as
well
\cite{Aslamazov_1959,Takada_1970,Agterberg_2001,Adachi_2003PRB,Vorontsov_2008PRB}.

Beside treating random impurities in an averaged fashion, some studies
treat impurities locally
\cite{Ting_2007PRB,YangPRB2008,Yanase_2009,Datta_2019PRB}, especially
in the case of single, few, or non-uniformly distributed
impurities. Such impurities, if strong enough, may lead to localized
states \cite{Vernier2011}.  When averaged over a large number of
uniform random distributions, it is expected that, for weak impurities,
these two approaches yield compatible results. Indeed, our result is
in agreement with the BdG based findings of Ref.~\cite{YangPRB2008}
for $s$-wave pairing, in that the FFLO state is much more sensitive to
disorders than the BCS state, and that it can survive moderate
disorder strength but may be fully suppressed by higher impurity
levels. Both results indicate that a low impurity level, or
equivalently a long mean free path, is needed for the FFLO state to
survive the disorder effect.

It should be noted that, BdG calculations are usually done in a
discretized lattice, which necessarily needs to be much larger than
the coherence length $\xi_0^{}$ of the superfluid, of the order of
$\hbar v_\textbf{F}^{}/\Delta$. In the BCS regime, the gap is small so
that $\xi_0^{}$ is huge. For a $d$-wave superconductor, due to the
nonlocal effect \cite{Kosztin_nonlocal}, $\xi_0^{}$ diverges in the
nodal directions. Both these cases raise a concern about the
quantitative reliability for BdG calculations, when the lattice size
is not big enough.


While we consider the ground state only, the treatment in principle
can be extended to finite temperatures at the mean-field
level. Without considering the FFLO state, the GMB effect acts
essentially as a reduction of the pairing interaction strength (with a
small temperature dependence) \cite{Qijin}. This would thus lead to a
reduction to both $T_\text{c}$ and $\Delta_0$, with a slight
difference between finite and zero $T$. (This difference vanishes in
the BCS limit). In this case, there should be a GMB-reduced pairing
temperature, $T^{*,\text{GMB}}$, at which pairs form but do not Bose
condense. Then, at a lower temperature, $T_\text{c}^\text{GMB}$, phase
coherence sets in and pairs start to Bose condense. The situation is
different with a nonzero FFLO wavevector $\mathbf{q}$, which is
pertinent to a high population imbalance or a high magnetic
field. While the FFLO mean-field solution usually exists at low $T$,
when pairing fluctuations, which usually lead to the formation of a
pseudogap, are taken into account, the mean-field FFLO states become
unstable, in the absence of extrinsic symmetry breaking factors such
as spatial anisotropy and lattices, as found in
Ref.~\cite{Insta}. Similar results were found by others as well
\cite{Marenko,Ashvin,RadzihovskyPRA84}. Even at the mean-field level,
there may exist an intermediate temperature pseudogap regime, between
$T^{*,\text{GMB}}$ and $T_\text{c}^\text{MF,GMB}$. An example of such a pseudogap regime,
calculated in the absence of the GMB corrections, can be found in
Ref.~\cite{Qijin}.

Finally, it is known that in the $H$--$T$ phase diagram of a superconductor,
the existence of the mean-field FFLO phase extends the $H_{\text{c,2}}$
line at low $T$ towards the high field side of its BCS counterpart,
leading to a kink-like feature at the tricritical point which signals
the onset of the FFLO state.  Since the field strength $H$ is proportionally
related to the population imbalance $p$, a counterpart $T$--$p$ phase
diagram can often be found in the atomic Fermi gas literature, e.g.,
Refs.~\cite{helianyi06prb,FFLO_MF_us}. Now that the GMB effect leads mainly to
a reduction of the pairing interaction strength, it is expected that
the $T$--$p$ phase diagram looks qualitatively similar to its clean counterpart at the reduced pairing strength.

\vskip 2ex

\section{Conclusion}
\label{conc}
 
In summary, we have investigated in homogeneous 3D systems the GMB
correction to the chemical potential difference $h/\Delta_0$, which is
responsible for the transition to the FFLO phase. We find at the
mean-field level that the window for the FFLO phase to exist has been
reduced by a factor of $(4e)^{-1/3}$. Therefore, the region in the
phase space that otherwise possesses an FFLO order will take
alternative solutions, such as phase separation and polaronic normal
state. This shall thus further confine the phase space where the true
stable solution is yet to be determined. 

We have also considered the GMB effect on the FFLO window in the
presence of weak (nonmagnetic) impurities or defects, in terms of a
finite lifetime $\tau = 1/\gamma$ of the quasi-particle
excitations. We find that a high impurity level leads to a reduction
in the critical field $h_\text{s}$ of the continuous phase transition
between the FFLO and the normal phase. This will shrink or completely
destroy the FFLO window.

\begin{acknowledgments} 

  H.~C. wish to thank CNPq and FAPEMIG for partial financial support,
  and Q.~C. is supported by NSF of China under grants No. 11774309 and
  No. 11674283. H. C. acknowledge discussions with M. A. R. Griffith.

\end{acknowledgments} 

\vspace{0.5cm}

\appendix

\section{Calculation of the pair susceptibility $\chi$ in the presence of impurities}
\label{ApA}

The dynamic pair susceptibility is given by,
\begin{eqnarray}
\label{chi1-3d}
\chi({\vec{q}}, \Omega)&=&
\sum_{\vec{k}} \frac{1-f(\xi_{\vec{k}-\vec{q}/2,\uparrow})-f(\xi_{\vec{k}+\vec{q}/2,\downarrow})}{\xi_{\vec{k}-\vec{q}/2,\uparrow}+\xi_{\vec{k}+\vec{q}/2,\downarrow}- \Omega}\\
\nonumber
                       &=&
                           \frac{1}{2} \sum_{\vec{k}} \frac{\tanh(\beta \xi_{\vec{k}-\vec{q}/2,\uparrow})+\tanh(\beta \xi_{\vec{k}+\vec{q}/2,\downarrow})}{\xi_{\vec{k}-\vec{q}/2,\uparrow}+\xi_{\vec{k}+\vec{q}/2,\downarrow}- \Omega}\,,
\end{eqnarray}
%
which can be rewritten as
\begin{equation}
\label{chi3-3d}
\chi({\vec{q}}, \Omega) =
\frac{1}{2} \sum_{\vec{k}} \left[ \frac{\tanh(\beta \xi_{\vec{k},\uparrow})}{\xi_{\vec{k},\uparrow}+\xi_{\vec{k}+\vec{q},\downarrow}- \Omega} + \frac{\tanh(\beta \xi_{\vec{k},\downarrow})}{\xi_{\vec{k}-\vec{q},\uparrow}+\xi_{\vec{k},\downarrow}- \Omega} \right]\,.
\end{equation}
The denominators can be approximated as
$\xi_{\vec{k},\uparrow}+\xi_{\vec{k}+\vec{q},\downarrow} - \Omega
\simeq 2(\xi_{\vec{k},\uparrow} + h + a \cos\theta- \Omega/2)$ and
$\xi_{\vec{k}-\vec{q},\uparrow}+\xi_{\vec{k},\downarrow} -
\Omega\simeq 2(\xi_{\vec{k},\downarrow} - h - a \cos\theta-
\Omega/2)$, where $a \equiv kq/2m$, and and $\theta$ is the angle between $\vec{k}$ and $\vec{q}$. Terms of order $q^2$ and
higher have been neglected. Then we obtain
\begin{eqnarray}
\label{chi4-3d}
\chi({\vec{q}}, \Omega)&=&
\frac{1}{4} \sum_{\vec{k}} \Bigg[\frac{\tanh(\beta \xi_{\vec{k},\uparrow})}{\xi_{\vec{k},\uparrow} + h + a \cos\theta- \Omega/2} \nonumber\\
&&{}+
\frac{\tanh(\beta \xi_{\vec{k},\downarrow})}{\xi_{\vec{k},\downarrow} - h - a \cos\theta- \Omega/2}\Bigg].
\end{eqnarray}
Now we first integrate out $\theta$ over a narrow momentum shell with
an energy cutoff $\omega_c$ near the Fermi level, followed by
analytical continuation, $\Omega \to \Omega + i\gamma$.  Then we
arrive in the static limit at 
\begin{eqnarray}
\label{chi9-g7-3d}
&\mathrm{Re}& \chi({\vec{q}}, \gamma)\nonumber\\ &&{} =  
\nonumber
 \frac{m^2}{4 \pi^2 q} \int_{0}^{\omega_c}\!\!\!\!  \mathrm{d} \omega\; \tanh\left(\!\frac{\beta \omega}{2}\right)\, \ln  \frac{(\omega + h + a)^2 +\gamma^2/4}{(\omega + h - a)^2 +\gamma^2/4} ,
\end{eqnarray}
where  $a \to v_\text{F}^{} q/2$, and $v_\text{F}^{}=k_\text{F}^{}/m$
is the Fermi velocity. Taking now the zero temperature limit, and
integrating over $\omega$ we obtain Eq.~(\ref{chi10-g8-3d}).

%

\end{document}